\documentclass[showpacs,amssymb,amsmath,aps,prd,secnumarabic,twocolumn,groupedaddress]{revtex4-1}

\usepackage{graphicx}

\newcommand{\jpsi}{J/$\psi$}

\newcommand{\chiAll}{$\chi$}

\newcommand{\chiOne}{$\chi_{1}$}
\newcommand{\chiTwo}{$\chi_{2}$}
\newcommand{\chicAll}{$\chi_c$}

\newcommand{\chicOne}{$\chi_{c1}$}
\newcommand{\chicTwo}{$\chi_{c2}$}
\newcommand{\upsAll}{$\Upsilon$}

\newcommand{\chibAll}{$\chi_b$}

\begin{document}

\title{
 Determination of $\chi_c$ and $\chi_b$ polarizations
from dilepton angular distributions in radiative decays}

\author{Pietro Faccioli$^a$, Carlos Louren\c{c}o$^b$, Jo\~ao
  Seixas$^{a,c}$, and Hermine K. W\"{o}hri$^{a,b}$}

\affiliation{$^a$Laborat\'orio de Instrumenta\c{c}\~ao e F\'{\i}sica
Experimental de Part\'{\i}culas (LIP), 1000-149 Lisbon, Portugal\\
$^b$European Organization for Nuclear Research (CERN), 1211 Geneva 23, Switzerland\\
$^c$Physics Department, Instituto Superior T\'ecnico (IST), 1049-001 Lisbon,
Portugal}

\date{\today}

\begin{abstract}

The angular distributions of the decay products in the successive decays
$\chi_{c} (\chi_{b}) \rightarrow \mathrm{J}/\psi (\Upsilon)\, \gamma$ and
$\mathrm{J}/\psi (\Upsilon) \rightarrow \ell^+ \ell^-$ are calculated
as a function of the angular momentum composition of the decaying \chiAll\
meson and of the multipole structure of the photon radiation, using a formalism
independent of production mechanisms and polarization frames.
The polarizations of the \chiAll\ states produced in high energy
collisions can be derived from the dilepton decay distributions of the daughter
\jpsi\ or \upsAll\ mesons, with a reduced dependence on the
details of the photon reconstruction or simulation.
Moreover, this method eliminates the dependence of the polarization measurement
on the actual details of the multipole structure of the radiative transition.
Problematic points in previous calculations of the \chicAll\ decay angular
distributions are identified and clarified.

\end{abstract}



\maketitle

\sloppy

\section{Introduction}
\label{sec:intro}

The existing \jpsi\ and \upsAll\ polarization measurements make no distinction
between directly produced states and those resulting from the decay of
higher-mass states. \jpsi\ and \upsAll\ mesons coming from \chiAll\ decays
have, in principle, very different polarizations with respect to the directly
produced ones. In fact, directly produced \chiAll\ and directly produced \jpsi\
or \upsAll\ have different angular momentum and parity properties, and
originate from different partonic processes. Moreover, the angular momentum
composition of the indirectly produced states is influenced by the presence of
the accompanying decay photon. Therefore, \chicAll\ and \chibAll\ polarization
measurements, together with the knowledge of how these states transmit their
polarizations when they decay, are essential in the understanding of the
observed \jpsi\ and \upsAll\ polarization patterns. An improved account of
feed-down effects in quarkonium polarization measurements, and calculations,
can shed new light in the interpretation of the significant discrepancy
existing today between the theory predictions and the experimental
data~\cite{bib:qqbarExpClar}.

In this paper we examine how the polarization is transmitted in the decays from
$P$ to $S$ quarkonium states. We study the angular distributions of the
successive decays
 $\chi_{c} \, (\chi_{b}) \rightarrow \mathrm{J}/\psi \, (\Upsilon)\, \gamma$ and
 $\mathrm{J}/\psi \, (\Upsilon) \rightarrow \ell^+ \ell^-$. We discuss
the sensitivity of these observable distributions to the angular momentum
composition (``polarization'') of the decaying \chiAll\ meson and their
additional dependence on the orbital angular momentum of the photon. As a
result of the study, we propose a convenient way of measuring \chiAll\
polarizations in high-energy experiments, essentially independent of the
details regarding the photon detection and of the magnitude of the higher-order
multipoles of the radiative transition.
This method is valid irrespectively of the production process (hadroproduction,
photoproduction, etc.).
The paper finishes with a critical review of previous calculations of the
\chicAll\ decay angular distributions, identifying and clarifying the causes
of their seemingly contradictory results.

\section{\boldmath Radiative decay amplitudes}
\label{sec:amplitude}

\begin{figure*}[t!]
\centering
\includegraphics[width=1.0\linewidth]{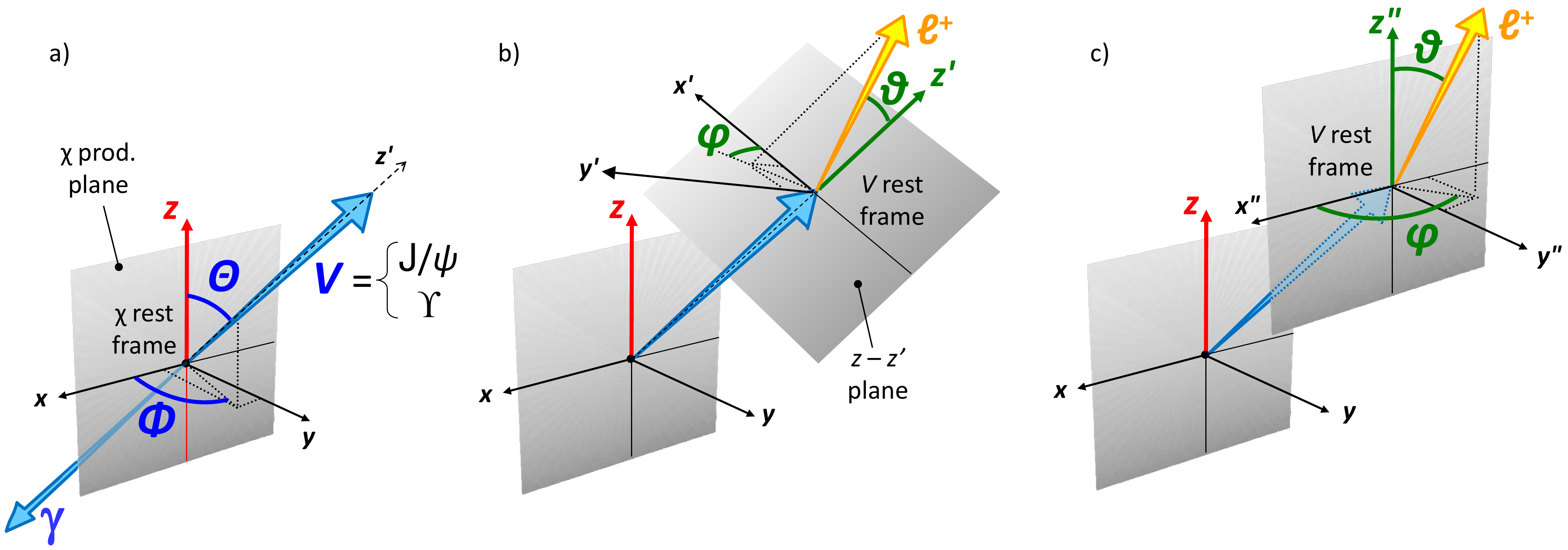}
\caption{Definition of axes and decay angles for $\chi \rightarrow V \gamma$
(a) and for $V \rightarrow \ell^+ \ell^-$ in two options, with the dilepton
polarization axis being the $V$ direction in the $\chi$ rest frame
(b) or parallel to the $\chi$ polarization axis (c).}
\label{fig:coords}
\end{figure*}

Throughout this paper, we generically denote by $V$ the charmonium and
bottomonium $^3\!S_1$ states, $\mathrm{J}/\psi$ and $\Upsilon$, and by \chiAll\
the $^3\!P_{j}$ states, $\chi_{cj}$ and $\chi_{bj}$, with $j=1,2$.
%
%
Without loss of generality for the discussions in this paper, we assume that the
$^3\!P_{j}$ state $\chi_j$ is produced in a single ``subprocess'' as a
given superposition of ${\rm J}_z$ eigenstates ($z$ being the
quantization axis chosen for the \chiAll\ angular momentum),
\begin{equation}
|\chi_j \rangle = \sum_{m=-j}^{j} b_m \, |\chi; \, j,m\rangle \, ,
\label{eq:chi_state}
\end{equation}
with ${\bf J}^2 |\chi; \, j,m\rangle = j(j+1) \, |\chi; \, j,m\rangle$ and
${\rm J}_z |\chi; \, j,m\rangle = m \, |\chi; \, j,m\rangle$. Notations for
axes and angles are shown in Fig.~\ref{fig:coords}(a). The total angular
momentum carried by the photon can have any (non-vanishing) value, while its
projection along the momentum direction of the $\gamma$ (and of $V$), the
$z^\prime$ axis, can only be $k^\prime = +1$ or $-1$, because the orbital
component has, by definition, zero projection along this direction. In other
words, the photon angular momentum state is an eigenstate of ${\rm
J}_{z^\prime}$ but, in general, neither of ${\bf J}^2$ nor of ${\rm J}_z$. It
can be represented as a complete expansion over eigenstates of ${\bf J}^2$ and
${\rm J}_z$, as
\begin{equation}
| \gamma ; \, k^\prime \rangle \; = \; \sum_{l = 1}^{\infty} \frac{\sqrt{2 l
+1}}{4 \pi} \sum_{k = - l}^{l} \mathcal{D}_{k \, k^\prime}^{l}(\Theta, \Phi) \,
| \gamma; \, l, k \rangle \, , \label{eq:photon_state}
\end{equation}
where ${\bf J}^2 | \gamma; \, l, k \rangle = l (l+1) \, | \gamma; \, l, k
\rangle$, ${\rm J}_z | \gamma; \, l, k \rangle = k \, | \gamma; l, k \rangle$
and the coefficients $\mathcal{D}_{k \, k^\prime}^{l}(\Theta,
\Phi)$~\cite{bib:BrinkSatchler} are the matrix elements of the rotation
corresponding to the change of quantization axis from $z^\prime$ (``natural''
quantization axis of the photon) to $z$ (\chiAll\ quantization axis adopted in
the measurement).

The amplitude of the radiative transition from the $\chi_j$ state to the $V$
state plus a photon having spin projection
$k^\prime$ along $z^\prime$ is
\begin{align}
\begin{split}
& \mathcal{A}(\chi_j \to V \gamma_{k^\prime}) \; \propto \; \sum_{m=-j}^{j}
b_m \sum_{l = 1}^{\infty} \sqrt{2 l +1}  \\
& \times \, \sum_{k = - l}^{l} \mathcal{D}_{k \, k^\prime}^{l
*}(\Theta, \Phi) \, \langle V \gamma_{k^\prime}; \, 1, m-k, l, k \, | \,
\mathcal{H} \, | \, \chi; \, j, m \rangle \, . \label{eq:decay_amplitude}
\end{split}
\end{align}
The matrix element of the elementary transition can be parametrized as
\begin{align}
\begin{split}
& \langle V \gamma_{k^\prime}; \, 1, m-k, l, k \, | \, \mathcal{H} \, | \, \chi;
\, j, m \rangle  \\
&  = \; (-1)^{\frac{1-k^\prime}{2}l} \, H_{jl} \; \langle 1, m-k, l, k \, | \,
j, m \rangle \, , \label{eq:transition_matrix}
\end{split}
\end{align}
where we have factored out the $k^\prime$-dependent sign, determined by
imposing that the photon distribution
\begin{equation}
W_{j}(\Theta, \Phi) \, = \, \sum_{k^\prime = \pm 1} |\mathcal{A}(\chi_j \to V
\gamma_{k^\prime})|^2 \label{eq:photon_distr}
\end{equation}
is parity invariant and using the property
\begin{equation}
\mathcal{D}_{k \, k^\prime}^{l}(\pi - \Theta, \pi + \Phi) = (-1)^{l-k^\prime}
e^{2ik^\prime \Phi} \, \mathcal{D}_{k\, -k^\prime}^{l}(\Theta, \Phi) \, .
\label{eq:parity_Dmatrix}
\end{equation}
The sums in Eq.~\ref{eq:decay_amplitude} only include terms in which the
Clebsch-Gordan coefficient $\langle 1, m-k, l, k \, | \, j, m \rangle$ is well
defined, i.e.\ when
\begin{equation}
1 \le l \le j + 1 \, , \quad |m-k| \le 1 \, . \label{eq:sum_indices}
\end{equation}

The partial amplitudes $H_{jl}$ denote $\chi_j \to V$ transitions with the
emission of a photon of total angular momentum $l$. In the spectroscopic
language they represent electric and magnetic $2^l$-pole radiations (dipole,
quadrupole, octupole, etc.), indicated with E$l$ and M$l$, respectively. The
two types of transitions differ in their parity properties: the electric
$2^l$-pole radiation has parity $(-1)^l$ while the magnetic $2^l$-pole
radiation has parity $(-1)^{l+1}$. Since the $^3P_j$ and $^3S_1$ quarkonium
states have opposite parities, the only allowed transitions are E1 (for all
\chiAll\ states), M2 (for \chiOne\ and \chiTwo) and E3 (for the \chiTwo).
Hereafter we use the short notations
\begin{align}
\begin{split}
h_l & = H_{1,l} \quad \rm{with} \quad l = 1,2 \, , \\
g_l & = H_{2,l} \quad \rm{with} \quad l = 1,2,3 \, ,
\end{split}
\end{align}
%
with the normalizations
\begin{equation}
h_1^2 + h_2^2 \; = \; g_1^2 + g_2^2 + g_3^2 \; = \; 1 \, .
\label{eq:multipoles_norm}
\end{equation}
In short, $h_1$ and $g_1$ represent, respectively for $\chi_1$ and $\chi_2$,
the relative amplitude of the E1 transition, $h_2$ and $g_2$ the corresponding
relative amplitudes of the M2 transition, and $g_3$ the relative amplitude of
the E3 transition (only for the $\chi_2$ case). The hierarchies $g_3 < g_2 <
g_1$ and $h_2 < h_1$ are expected. In fact, in the generic expansion of the
radiation field around a system of oscillating charges in terms of angular
momentum eigenfunctions, the $l$-th term vanishes more and more rapidly, at
large distance from the origin, as $l$ increases. This behaviour reflects the
fact that the wavelength of the emitted photon, $\lambda_\gamma = hc/E_\gamma
\simeq hc/(0.4\,\mathrm{GeV}) \simeq 3$~fm, is sizeable with respect to the
dimensions of the quarkonium ($\simeq 0.4-0.7$~fm), so that at the typical
distance scale $r = \lambda_\gamma$ the electromagnetic field is already only
weakly sensitive to the internal charge and current distributions of the
radiating object. With respect to the first non-vanishing term, higher
multipole terms, produced by more complex charge/current configurations, are
therefore foreseen to be
\begin{table}[t]
\begin{tabular}{l@{\quad}c@{\quad}c@{\quad}c}
\toprule
Experiment                 &        $h_2$ [\%]      &      $g_2$ [\%]           &     $g_3$ [\%]           \\
\colrule
Crystal Ball~\cite{bib:CB} & $-0.2^{+0.8}_{-2.0}$  & $-33^{+11}_{-30}$        &          --             \\
E760~\cite{bib:E760}       &          --           & $-14 \pm 6$              &  $0^{+6}_{-5}$          \\
E835~\cite{bib:E835}       & $0.2 \pm 3.2 $ &$-9.3^{+3.9}_{-4.1}$&$2.0^{+5.6}_{-4.5}$\\
CLEO~\cite{bib:CLEO}       & $-6.26\pm0.67$ & $-9.3\pm1.6$ & $1.7\pm1.4$ \\
\botrule
\end{tabular}
\caption{Higher-order photon multipoles in $\chi_{c(1,2)}
\to {\rm J}/\psi \, \gamma$ decays. } \label{tab:multipoles}
\end{table}
increasingly smaller, even if not necessarily as suppressed as in nuclear
$\gamma$-ray transitions, where the emitted radiation has a wavelength several
orders of magnitude larger than the nuclear dimensions. The study of the $c
\bar{c}$ radiation multipoles addresses aspects of the quark model, including
the properties of the bound-state wave functions and the electro-magnetic
properties of the charm quark. For example, the relative contribution of the M2
amplitudes is significantly dependent on the corrections to the
charm
quark magnetic moment $\mu_c = \frac{2}{3}
\frac{e}{2m_c}$~\cite{bib:multipole_calculations}.
The existing $h_2$, $g_2$ and $g_3$
measurements, for the  \chicAll, are shown in Table~\ref{tab:multipoles}.
The M2 amplitude contribution is of order $10\%$ for the \chicTwo\ and even smaller
for the \chicOne, although the two most precise \chicOne\ results are
incompatible with each other.
No experimental information exists for \chibAll\ decays. We will
discuss in Sect.~\ref{sec:discussion} the effects induced by the higher-order
multipole contributions on the observable angular distributions.

\section{Photon distribution}
\label{sec:photdistr}

The angular distribution of the photon direction in the \chiAll\ rest frame, as
a function of the \chiAll\ angular momentum composition $\{b_m\}$ and of the
photon multipole amplitudes, is obtained by expanding
Eq.~\ref{eq:photon_distr}. The resulting $\chi_0$ distribution is spherically
symmetric, reflecting the rotational invariance of the $j=0$ angular momentum
state and the imposed parity invariance of the decay. As for the \chiOne\
decay, the expression of the angular distribution is
\begin{align} \begin{split} \label{eq:photon_distr_chi1}
& W_{1}(\Theta, \Phi) \, = \, \frac{3}{4 \pi (3+\lambda_\Theta)} \, (1 \, + \,
\lambda_\Theta \, \cos^2\!\Theta  \\
& +  \lambda_\Phi \,
\sin^2\!\Theta \cos2\Phi \, + \, \lambda_{\Theta \Phi} \, \sin2\Theta \, \cos\Phi \\
& +  \lambda^{\perp}_\Phi \, \sin^2\!\Theta \sin2\Phi \, + \,
\lambda^{\perp}_{\Theta \Phi} \, \sin2\Theta \, \sin\Phi ) \, ,
\end{split} \end{align}
where
\begin{align} \begin{split} \label{eq:photon_distr_chi1_coeffs}
\lambda_\Theta   & =  \frac{1}{D}(1 - 3\Delta ) [ {2 |b_0|^2 - ( |b_{+1}|^2 +
|b_{-1}|^2 )}] \, ,  \\
\lambda_\Phi  & =   - \frac{2} {D}(1 - 3\Delta )\operatorname{Re}
(b_{ + 1}^ *  b_{ - 1} ) \, ,  \\
\lambda_{\Theta \Phi }  & =   - \frac{{\sqrt 2 }} {D}(1 - 3\Delta
)\operatorname{Re} [b_0^*  (b_{ + 1} - b_{ - 1} )] \, ,  \\
\lambda_\Phi^{\perp}  & =   - \frac{2} {D}(1 - 3\Delta )\operatorname{Im}
(b_{ + 1}^ *  b_{ - 1} ) \, ,  \\
\lambda_{\Theta \Phi }^{\perp}  & =   \frac{{\sqrt 2 }} {D}(1 - 3\Delta
)\operatorname{Im} [b_0^*  (b_{ + 1} + b_{ - 1} )] \, ,
\end{split} \end{align}
with
\begin{align} \begin{split}
D & =  2(1 + \Delta )\left| {b_0 } \right|^2  + (3 - \Delta )(\left| {b_{ + 1}
} \right|^2 + \left| {b_{ - 1} } \right|^2 ) \, ,  \\
\Delta  & =  - 2\,h_1\,h_2 \, .
\end{split} \end{align}
\begin{figure*}[ht]
\centering
\includegraphics[width=0.85\linewidth]{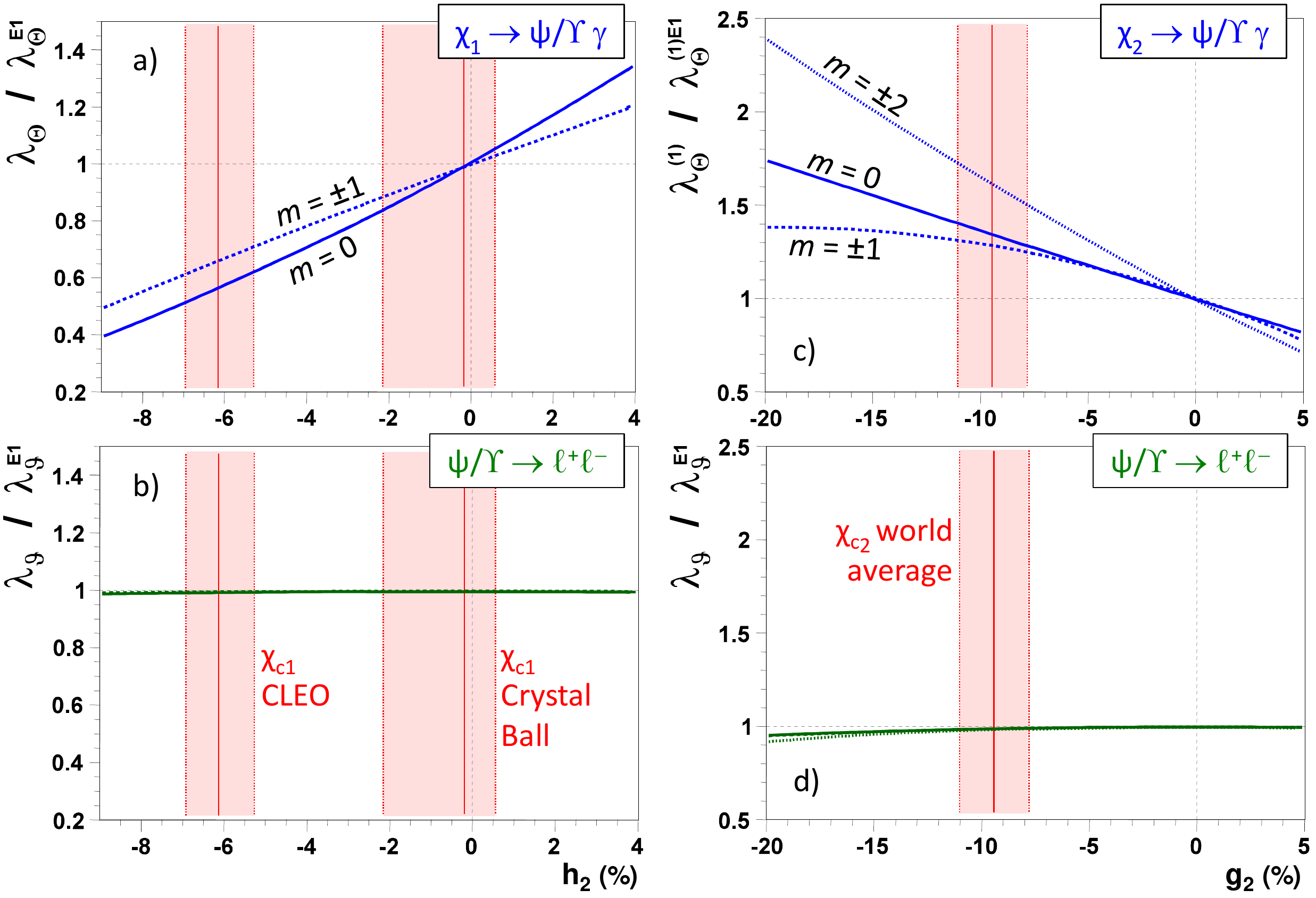}
\caption{Dependence of the parameters $\lambda_\Theta$ (a) and
$\lambda_\vartheta$ (b) of the $\chi_{1}$ photon and dilepton distributions and
of the parameters $\lambda_\Theta^{(1)}$ (c) and $\lambda_\vartheta$ (d) of the
$\chi_{2}$ photon and dilepton distributions on the relative contribution of
the magnetic quadrupole transition amplitude. } \label{fig:multipoles}
\end{figure*}
The angular distribution of the photon from the \chiTwo\ decay,
significantly more complex, is
%
\begin{align} \begin{split}
& W_{2}(\Theta, \Phi) = \frac{15}{4 \pi (15 + 5 \lambda^{(1)}_\Theta + 3
\lambda^{(2)}_\Theta)} \, ( 1 \, + \, \lambda^{(1)}_\Theta \, \cos^2\!\Theta \, \\
& + \, \lambda^{(2)}_\Theta \, \cos^4\!\Theta \,
  + \, \lambda^{(1)}_\Phi \, \sin^2\!\Theta \cos2\Phi \,
  + \, \lambda^{(2)}_\Phi \, \sin^4\!\Theta \cos2\Phi \, \\
& + \, \lambda^{(3)}_\Phi \, \sin^4\!\Theta \cos4\Phi  \,
 + \, \lambda^{(1)\perp}_\Phi \, \sin^2\!\Theta \sin2\Phi \, \\
& + \, \lambda^{(2)\perp}_\Phi \, \sin^4\!\Theta \sin2\Phi \,
 + \, \lambda^{(3)\perp}_\Phi \, \sin^4\!\Theta \sin4\Phi  \\
& + \, \lambda^{(1)}_{\Theta \Phi} \, \sin2\Theta \cos\Phi \,
 + \, \lambda^{(2)}_{\Theta \Phi} \, \sin^2\!\Theta \sin2\Theta \cos\Phi \, \\
&+ \, \lambda^{(3)}_{\Theta \Phi} \, \sin^2\!\Theta \sin2\Theta \cos3\Phi \,
 + \, \lambda^{(1)\perp}_{\Theta \Phi} \, \sin2\Theta \sin\Phi \, \\
& + \, \lambda^{(2)\perp}_{\Theta \Phi} \, \sin^2\!\Theta \sin2\Theta \sin\Phi \,
+ \, \lambda^{(3)\perp}_{\Theta \Phi} \, \sin^2\!\Theta \sin2\Theta \sin3\Phi ) \, ,
\end{split} \end{align}
%
where
\begin{align} \begin{split}
\label{eq:photon_distr_chi2_coeffs}
 \lambda_\Theta^{(1)} &=  -\frac{3}{D} [ 2(1 + \Delta_1 ) |{b_0}| ^2 \\
& + (1 - \tfrac{2}{3} \Delta_1  - \tfrac{5}{3} \Delta_2 )( |b_{+ 1}|^2 + |b_{ - 1}|^2 ) \\
& - (2 + \tfrac{1}{3} \Delta_1 - \tfrac{5}{3}\Delta_2) (|b_{+2}|^2 + |b_{-2}|^2) ] \, , \\
 \lambda_{\Theta}^{(2)} &= \frac{\Delta}{2} [ 6 |b_0|^2 - 4(|b_{ + 1}|^2 +|{b_{-1}|^2 ) + |b_{ + 2}|^2 + |b_{-2}|^2 } ] \, , \\
 \lambda_\Phi^{(1)} &=  \frac{2}{D} \operatorname{Re}[\sqrt 6 (1 + \Delta_1 )
\, b_0^* (b_{+2} + b_{-2}) \\
& + (3 - 2\Delta_1 - 5\Delta_2 ) \, b_{+1}^* b_{-1} ] \, , \\
 \lambda_\Phi^{(2)} &=  - \Delta \operatorname{Re} [
\sqrt 6 \, b_0^* (b_{+2} + b_{-2}) - 4 \, b_{+1}^* b_{-1} ] \, , \\
 \lambda_\Phi^{(3)} &=  \Delta \operatorname{Re}(b_{+2}^* b_{-2}) \, , \\
 \lambda_\Phi^{(1)\perp} &=  \frac{2}{D} \operatorname{Im}[-\sqrt 6 (1 +
\Delta_1 ) \, b_0^* (b_{+2} - b_{-2}) \\
& + (3 - 2\Delta_1 - 5\Delta_2 ) \, b_{+1}^* b_{-1} ] \, , \\
 \lambda_\Phi^{(2)\perp} &=  \Delta
\operatorname{Im} [ \sqrt 6 \, b_0^* (b_{+2} - b_{-2}) + 4 \, b_{+1}^* b_{-1} ]\, , \\
 \lambda_\Phi^{(3)\perp}  &=  \Delta
\operatorname{Im} (b_{+2}^* b_{-2}) \, , \\
 \lambda_{\Theta \Phi}^{(1)} &=  \frac{1}{D} \operatorname{Re}[\sqrt 6 (1 -
\tfrac{2}{3}\Delta_1  - \tfrac{5}{3} \Delta_2 ) \, b_0^*(b_{+1}-b_{-1})  \\
& +  (6 + \tfrac{8}{3} \Delta_1 - \tfrac{10}{3} \Delta_2 ) \, (b_{+2}^* b_{+1}
- b_{-2}^ * b_{-1} )] \, , \\
 \lambda_{\Theta \Phi}^{(2)} &=  \Delta
\operatorname{Re} [\sqrt 6 \, b_0^* (b_{+1}-b_{-1}) - \, (b_{+2}^* b_{+1} -
b_{-2}^* b_{-1} )] \, , \\
 \lambda_{\Theta \Phi}^{(3)} &= \Delta
\operatorname{Re} (b_{+2}^* b_{-1}- b_{-2}^* b_{+1} ) \, , \\
 \lambda_{\Theta \Phi}^{(1)\perp} &=  \frac{1}{D} \operatorname{Im}[-\sqrt 6 (1
- \tfrac{2}{3}\Delta_1  - \tfrac{5}{3} \Delta_2 ) \, b_0^* (b_{+1}+b_{-1}) \\
& + (6 + \tfrac{8}{3} \Delta_1 - \tfrac{10}{3} \Delta_2 ) \, (b_{+2}^* b_{+1}
+ b_{-2}^ * b_{-1} )] \, , \\
 \lambda_{\Theta \Phi}^{(2)\perp} &=  -\Delta
\operatorname{Im} [\sqrt 6 \, b_0^* (b_{+1}+b_{-1}) + \, b_{+2}^* b_{+1} +
b_{-2}^* b_{-1} ] \, , \\
 \lambda_{\Theta \Phi}^{(3)\perp} &= \Delta
\operatorname{Im} (b_{+2}^* b_{-1}+ b_{-2}^*b_{+1} ) \, ,
\end{split} \end{align}
with
\begin{align} \begin{split}
D & =  (10 + \Delta_1 - \Delta_2) |b_0|^2  + (9 - \Delta_2) (|b_{ + 1}|^2 + |b_{-1}|^2) \\
&+  (6 - \tfrac{1}{2} \Delta_1 + \tfrac{3}{2} \Delta_2)(|b_{+2}|^2 + |b_{-2}|^2 ) \, , \\
\Delta_1  & =  4\,g_2^2  + 6 \sqrt 5 \, g_2 g_3 - 2 \sqrt 5 \, g_1 g_2 - 2
g_3^2 + 14 \, g_1 g_3 \, , \\
\Delta_2  & =  4\,g_2^2  + 4 \sqrt 5 \, g_2 g_3 + 2 \sqrt 5 \, g_1 g_2 + 3
g_3^2 + 4 \, g_1 g_3 \, , \\
\Delta & = 5/(3D) \, (\Delta_1 + \Delta_2) \, .
\end{split} \end{align}
%

As shown in Fig.~\ref{fig:multipoles}, the dependence of the photon
distribution on the \chiAll\ angular momentum configuration is very sensitive
to the contribution of the higher photon multipoles.
Figure~\ref{fig:multipoles}(c) shows, in particular, that the polar anisotropy
parameter $\lambda_\Theta^{(1)}$, at the average value of $g_2$ measured for
the \chicTwo\ (assuming $g_3=0$), is 30\% higher than the value expected in the
E1-dominance case if the \chiTwo\ polarization state is $m=\pm1$ or 70\% higher
if $m=\pm2$. This shows that the derivation of the average polarization state
in which the \chiAll\ is produced from the observed photon angular distribution
relies crucially on the knowledge of the multipole amplitudes. Seen from the
opposite perspective, we see that the so-called E1 approximation
($h_2=g_2=g_3=0$) is clearly not applicable in the calculation of the $\chi \to
V \gamma$ decay kinematics expected for a given \chiAll\ production mechanism.

\section{Lepton distribution}
\label{sec:lepdistr}

In the parity-conserving case here considered, the general expression for the
angular distribution of the dilepton decay of a vector state
is~\cite{bib:LTviolation}
%
\begin{align} \begin{split} \label{eq:dilepton_distr}
&w(\vartheta, \varphi) \, = \, \frac{3}{4 \pi (3+\lambda_\vartheta)} \, (1 \,
 + \, \lambda_\vartheta \, \cos^2\!\vartheta \\
& + \, \lambda_\varphi \, \sin^2\!\vartheta \, \cos2\varphi \,  + \,
\lambda_{\vartheta \varphi} \, \sin2\vartheta \, \cos\varphi \\
& + \, \lambda^{\perp}_\varphi \, \sin^2\!\vartheta \sin2\varphi \,
 + \, \lambda^{\perp}_{\vartheta \varphi} \, \sin2\vartheta \, \sin\varphi ) \,
 ,
\end{split} \end{align}
%
analogous in form to Eq.~\ref{eq:photon_distr_chi1}. The traditional choice of
axes, adopted in
calculations~\cite{bib:Olsson_Suchyta,bib:Ridener_Sebastian_Grotch} and
measurements~\cite{bib:CB,bib:E760,bib:E835,bib:CLEO} of the full decay angular
distribution for \chicAll\ mesons produced at low laboratory momentum, is
represented in Fig.~\ref{fig:coords}(b), where the $V$ polarization axis,
$z^\prime$, is the $V$ direction in the \chiAll\ rest frame. With respect to
this system of axes, any measurement will always find, for instance in the case
of the polar anisotropies for \chiOne\ and \chiTwo\ dileptons (neglecting, for
simplicity, the E3 contribution in the latter case), the following values:
\begin{align} \begin{split}
\lambda_\vartheta^{j=1} & = -\frac{1}{3} \left[ 1 - \frac{16}{3} h_2 +
\mathcal{O}(h_2^2) \right] \, , \\
\lambda_\vartheta^{j=2} & = \frac{1}{13} \left[ 1 - \frac{80 \sqrt{5}}{13} g_2
+ \mathcal{O}(g_2^2) \right] \, .
\end{split} \end{align}
The dilepton distribution in the $x^{\prime},y^{\prime},z^{\prime}$ coordinate
system is independent on the \chiAll\ polarization state. This choice of axes,
while suitable for measuring the contribution of the higher-order multipoles,
does not provide any information on the polarization of the \chiAll\ and,
hence, on its production mechanism.

We propose here an alternative definition of the $V$ polarization
frame, enabling the determination of the \chiAll\ polarization in high-momentum
experiments without the need of measuring the full photon-dilepton kinematic
correlations. This definition, shown in Fig.~\ref{fig:coords}(c), ``clones''
the \chiAll\ polarization frame, defined in the \chiAll\ rest frame, into the
$V$ rest frame, taking the $x^{\prime\prime},y^{\prime\prime},z^{\prime\prime}$
axes to be parallel to the $x,y,z$ axes.

The coefficients of the dilepton distribution can be written as a function of
the angular momentum composition of the decaying vector
state~\cite{bib:LTviolation}, $|V \rangle =
\sum_{n=-1}^{n=+1} a_n \, |V; \, 1,n\rangle$, as
\begin{align} \begin{split} \label{eq:lambdas_vs_amplitudes_1S}
  \lambda_{\vartheta} & =  \frac{{\mathcal{N}}-3 |a_0|^2}{\mathcal{N}+|a_0|^2}  \,
  , \quad
  \lambda_{\varphi}   =  \frac{ 2 \, \mathrm{Re} [a_{+1}^{*}
    a_{-1}] }{\mathcal{N}+|a_0|^2} \, ,  \\
  \lambda_{\vartheta \varphi} &  =  \frac{ \sqrt{2} \, \mathrm{Re} [ a_{0}^{*} ( a_{+1} - a_{-1})] }{\mathcal{N}+|a_0|^2} \, , \\
  \lambda^{\bot}_{\varphi}  & =  \frac{ 2 \, \mathrm{Im} [a_{+1}^{*} a_{-1}] }{\mathcal{N}+|a_0|^2} \, , \\
  \lambda^{\bot}_{\vartheta \varphi}  & = \frac{ - \sqrt{2} \,
    \mathrm{Im} [a_{0}^{*} (a_{+1} + a_{-1})] }{\mathcal{N}+|a_0|^2} \, ,
\end{split} \end{align}
where $\mathcal{N} = |a_0|^2 + |a_{+1}|^2 + |a_{-1}|^2$. The partial amplitude
of the $\chi_j$ decay into a vector state with angular momentum projection $n$
($=-1,0,1$) along $z^{\prime\prime}$ and a photon with angular momentum
projection $k^\prime$ ($=-1,1$) along $z^{\prime}$, from
Eqs.~\ref{eq:decay_amplitude} and~\ref{eq:transition_matrix}, is
%
\begin{align} \begin{split}
&
a_n^{(j,k^\prime)}(\Theta,\Phi) \; \propto \; \sum_{m=-j}^{j}
\sum_{l =1}^{j+1} \sum_{k = - l}^{l} \delta_{m-k,n} \; b_m  \sqrt{2 l +1} \\
&
\times \mathcal{D}_{k \, k^\prime}^{l*}(\Theta, \Phi) \, (-1)^{\frac{1-k^\prime}{2}l}
\, H_{jl} \; \langle 1, m-k, l, k \, | \, j, m \rangle  \, .
\label{eq:V_partial_amplitude}
\end{split} \end{align}
%
Inserting these amplitudes into the expressions of the coefficients in
Eq.~\ref{eq:lambdas_vs_amplitudes_1S} and averaging over the photon states
$k^\prime= \pm 1$ according to the sum rule
\begin{equation}
X \, = \, \frac{ \frac{ N^{(k^\prime=+1)} \, X^{(k^\prime=+1)} } { 3 +
\lambda_{\vartheta}^{(k^\prime=+1)} } + \frac { N^{(k^\prime=-1)} \,
X^{(k^\prime=-1)} }{ 3 + \lambda_{\vartheta}^{(k^\prime=-1)} } } { \frac{
N^{(k^\prime=+1)} }{ 3 + \lambda_{\vartheta}^{(k^\prime=+1)} } + \frac{
N^{(k^\prime=-1)} }{ 3 + \lambda_{\vartheta}^{(k^\prime=-1)} } } \, ,
\label{eq:kappa_average}
\end{equation}
with $X = \lambda_{\vartheta}$, $\lambda_{\varphi}$, etc. and
$N^{(k^\prime=-1)} = N^{(k^\prime=+1)}$ for parity conservation, it is possible
to obtain the expression of the full angular distribution
$W(\Theta,\Phi,\vartheta,\varphi)$ of the decay process $\chi \rightarrow V\,
\gamma \rightarrow V \ell^+ \ell^-$. In the following discussion, however, we
only consider the dilepton distribution, obtained by integrating
$W(\Theta,\Phi,\vartheta,\varphi)$ over $\Theta$ and $\Phi$.

In the frame defined in Fig.~\ref{fig:coords}(c), the dilepton decay
distribution of $V$
mesons originating from $\chi_0$ decays is isotropic.
In what concerns the state $|\chi_1 \rangle = \sum_{m=-1}^{m=+1} b_m \, |\chi;
\, 1,m\rangle$, the coefficients of the dilepton angular distribution are:
\begin{align} \begin{split}
\label{eq:dilepton_distr_chi1_coeffs}
\lambda_\vartheta   & =  \frac{1}{D_1} [ {2 |b_0|^2 - ( |b_{+1}|^2
+
|b_{-1}|^2 )}] \, ,  \\
\lambda_\phi  & =   - \frac{2} {D_1} \operatorname{Re}
(b_{ + 1}^ *  b_{ - 1} ) \, ,  \\
\lambda_{\vartheta \phi }  & =   - \frac{{\sqrt 2 }} {D_1}
\operatorname{Re} [b_0^*  (b_{ + 1} - b_{ - 1} )] \, ,  \\
\lambda_\phi^{\perp}  & =   - \frac{2} {D_1}\operatorname{Im}
(b_{ + 1}^ *  b_{ - 1} ) \, ,  \\
\lambda_{\vartheta \phi }^{\perp}  & =   \frac{{\sqrt 2 }} {D_1}
\operatorname{Im} [b_0^*  (b_{ + 1} + b_{ - 1} )] \, ,
\end{split} \end{align}
with
\begin{align} \begin{split}
D_1 & = D / (1 - 3\delta)\, , \\
D & =  2(1 + \delta )\left| {b_0 } \right|^2  + (3 - \delta )(\left| {b_{ + 1}
} \right|^2 + \left| {b_{ - 1} } \right|^2 ) \, ,  \\
\delta  & = \tfrac{2}{5} h_2^2 \, .
\end{split} \end{align}

The corresponding coefficients for the decay of the state $|\chi_2 \rangle =
\sum_{m=-2}^{m=+2} b_m \, |\chi; \, 2,m\rangle$ are
%
%
\begin{align} \begin{split} \label{eq:dilepton_distr_chi2_coeffs}
\lambda_\vartheta & = -\frac{3}{D_2} [ 2 |{b_0}| ^2 + |b_{+ 1}|^2 +
|b_{ - 1}|^2 - 2 (|b_{+2}|^2 + |b_{-2}|^2) ] \, ,  \\
\lambda_\varphi & = \frac{2}{D_2} \operatorname{Re}[\sqrt 6 \,
b_0^* (b_{+2} + b_{-2}) + 3 \, b_{+1}^* b_{-1} ] \, , \\
\lambda_{\vartheta \varphi} & =  \frac{1}{D_2}
\operatorname{Re}[\sqrt 6 \, b_0^*(b_{+1}-b_{-1}) + 6 \, (b_{+2}^* b_{+1} -
b_{-2}^ * b_{-1} )] \, , \\
\lambda_\varphi^{\bot} & = \frac{2}{D_2} \operatorname{Im}[-\sqrt 6
\,
b_0^* (b_{+2} - b_{-2}) + 3 \, b_{+1}^* b_{-1} ] \, , \\
\lambda_{\vartheta \varphi}^{\bot} & =  \frac{1}{D_2}
\operatorname{Im}[- \sqrt 6 \, b_0^*(b_{+1}+b_{-1}) + 6 \, (b_{+2}^* b_{+1} +
b_{-2}^* b_{-1} )] \, ,
\end{split} \end{align}
with
\begin{align} \begin{split}
D_2 & = D / (1 - \delta ) \, , \\
D & =  2 (5 - \delta) |b_0|^2  + (9 - \delta) (|b_{ + 1}|^2 + |b_{-1}|^2) \\
& +  2 (3 + \delta)(|b_{+2}|^2 + |b_{-2}|^2 ) \, , \\
\delta & =  2\,g_2^2  + \tfrac{5}{7} g_3^2 \, .
\end{split} \end{align}
%
%

Without experimental separation between the \chiOne\ and \chiTwo\ signals (the
$\chi_0 \rightarrow V \gamma$ contribution is negligible), the dilepton distribution
measurement effectively yields the corresponding average polarization
parameters, implicitly weighted by $N^{(j=1)}$ and $N^{(j=2)}$, respectively the numbers of
reconstructed dileptons coming from \chiOne\ and \chiTwo\ decays ($X =
\lambda_{\vartheta}$, $\lambda_{\varphi}$, etc.):
\begin{equation}
X \, = \, \frac{ \frac{ N^{(j=1)} \, X^{(j=1)}  } { 3 +
\lambda_{\vartheta}^{(j=1)} } + \frac { N^{(j=2)} \, X^{(j=2)} }{ 3 +
\lambda_{\vartheta}^{(j=2)} } } { \frac{ N^{(j=1)} }{ 3 +
\lambda_{\vartheta}^{(j=1)} } + \frac{ N^{(j=2)} }{ 3 +
\lambda_{\vartheta}^{(j=2)} } } \, . \label{eq:chi12average}
\end{equation}
%

\section{\boldmath Measurement of \chiAll\ polarization at high momentum }
\label{sec:discussion}

The formulas obtained in the previous two sections suggest two remarks. First,
with the choice of the $x^{\prime\prime},y^{\prime\prime},z^{\prime\prime}$
axes, the dilepton distribution contains as much information as the photon
distribution regarding the \chiAll\ polarization state. The two distributions
are even \emph{identical} when higher-order multipoles are neglected, as can be
recognized by comparing Eq.~\ref{eq:dilepton_distr_chi1_coeffs} with
Eq.~\ref{eq:photon_distr_chi1_coeffs} and
Eq.~\ref{eq:dilepton_distr_chi2_coeffs} with
Eq.~\ref{eq:photon_distr_chi2_coeffs} for $h_2=g_2=g_3=0$. In this limit, for
example, $\lambda_\vartheta = \lambda_{\Theta} = -1/3$ and $+1$, respectively
for pure $|j, m\rangle = |1,\pm 1\rangle$ and $|1,0\rangle$ \chiAll\ states,
and $\lambda_\vartheta = \lambda_{\Theta}^{(1)} = +1, -1/3$ and $-3/5$,
respectively for pure $|2,\pm 2\rangle$, $|2,\pm 1\rangle$ and $|2,0\rangle$
states, while the additional terms of the photon distribution in the \chiTwo\
case ($\lambda_{\Theta}^{(2)}$, $\lambda_\Phi^{(2)}$, $\lambda_\Phi^{(3)}$,
$\lambda_{\Theta \Phi}^{(2)}$ and $\lambda_{\Theta \Phi}^{(3)}$) vanish.
Second, the dependence of the dilepton distribution on the higher photon
multipoles is negligible, as shown in Fig.~\ref{fig:multipoles}(b,d) for
$\lambda_\vartheta$.

The definition of the $x,y,z$ axes (and, therefore, of the
$x^{\prime\prime},y^{\prime\prime},z^{\prime\prime}$ axes) uses the momenta of
the colliding hadrons \emph{as seen in the \chiAll\ rest frame}, so that it
requires, in general, the knowledge of the photon momentum. However, \emph{for
sufficiently high (total) momentum of the dilepton}, the \chiAll\ and $V$ rest
frames coincide and the $x^{\prime\prime},y^{\prime\prime},z^{\prime\prime}$
axes can be approximately defined using only momenta seen in the
$V$ rest frame. For example, if the \chiAll\ polarization axis ($z$) is defined
along the bisector of the beam momenta in the \chiAll\ rest frame
(Collins--Soper frame~\cite{bib:coll_sop}), the corresponding
$z^{\prime\prime}$ axis is approximated by the bisector of the beam momenta in
the \jpsi~/~\upsAll\ rest frame.
The relative error induced by this approximation on the polar anisotropy
parameter is
\begin{equation}
\left| \frac{\Delta\lambda_\vartheta}{\lambda_\vartheta} \right| =
\mathcal{O}\left[\left( \frac{\Delta M}{p} \right)^2 \right] \, ,
\label{eq:approx}
\end{equation}
where $\Delta M$ is the $\chi - V$
mass difference and $p$ is the
\emph{total} laboratory momentum of the dilepton. Therefore, for not-too-small
momentum the frame definition we propose coincides with the frame defined in
the measurement of the polarization of inclusively produced \jpsi~/~\upsAll\
mesons (Collins--Soper or helicity, for example). In other words, the
measurement of the dilepton distribution at sufficiently high laboratory
momentum provides a direct determination of the \chiAll\ polarization along the
chosen polarization axis. This determination is cleaner than the one using the
photon distribution in the \chiAll\ rest frame, because it is independent of
the knowledge of the higher-order photon multipoles.

The above-mentioned approximation, in which the system of axes
$x^{\prime\prime},y^{\prime\prime},z^{\prime\prime}$ is set without any
knowledge of the photon momentum, becomes rapidly invalid as $p \to 0$. The
$\chi_{0}$ case, although of little practical importance in the scope of this
paper (the branching ratio of the $\chi_{c0}$ decay to $\mathrm{J}/\psi$, for
example, is only $\simeq 1\%$), can be used to give a simple illustration of
what happens going from high to low $p$. As discussed above, at high $p$ the
polarization of dileptons from $\chi_{0}$ vanishes in the helicity frame (as
well as in any other frame defined ignoring the photon momentum), mirroring the
perfect isotropy of the photon emission in the $\chi_{0}$ rest frame. On the
other hand, the $1S$ state coming from the $j=0$ \chiAll\ state has an
intrinsic spin alignment, always opposite to the one of the photon
($\textbf{J}_{V} + \textbf{J}_{\gamma} = \textbf{J}_{\chi_{0}} = \textbf{0}$);
in other words, a fully transverse $1S$ polarization is observed if the
direction of the photon in the $\chi_{0}$ rest frame [$z^\prime$ in
Fig.~\ref{fig:coords}(b)] is taken as reference axis. In the low-$p$ limit,
when the $\chi_{0}$ tends to be produced at rest in the laboratory, that
direction tends to coincide with the center-of-mass helicity axis. In short, if
we choose the center-of-mass helicity frame, the $V$ polarization equals the
zero polarization of the $\chi_0$ only at high momentum, while it changes to
fully transverse at low momentum, where it simply reflects the intrinsic photon
polarization. This example shows that the possibility to measure the \chiAll\
polarization from the dilepton distribution ignoring the photon momentum is
strictly limited to a kinematic domain where $p \gg \Delta M$. However, the
error in Eq.~\ref{eq:approx} is already as small as 1\% when $p>4$~GeV/$c$,
a condition fulfilled, in particular, by essentially all the quarkonium
events collected by the LHC experiments.

\begin{figure}[t!]
\centering
\includegraphics[width=1.0\linewidth]{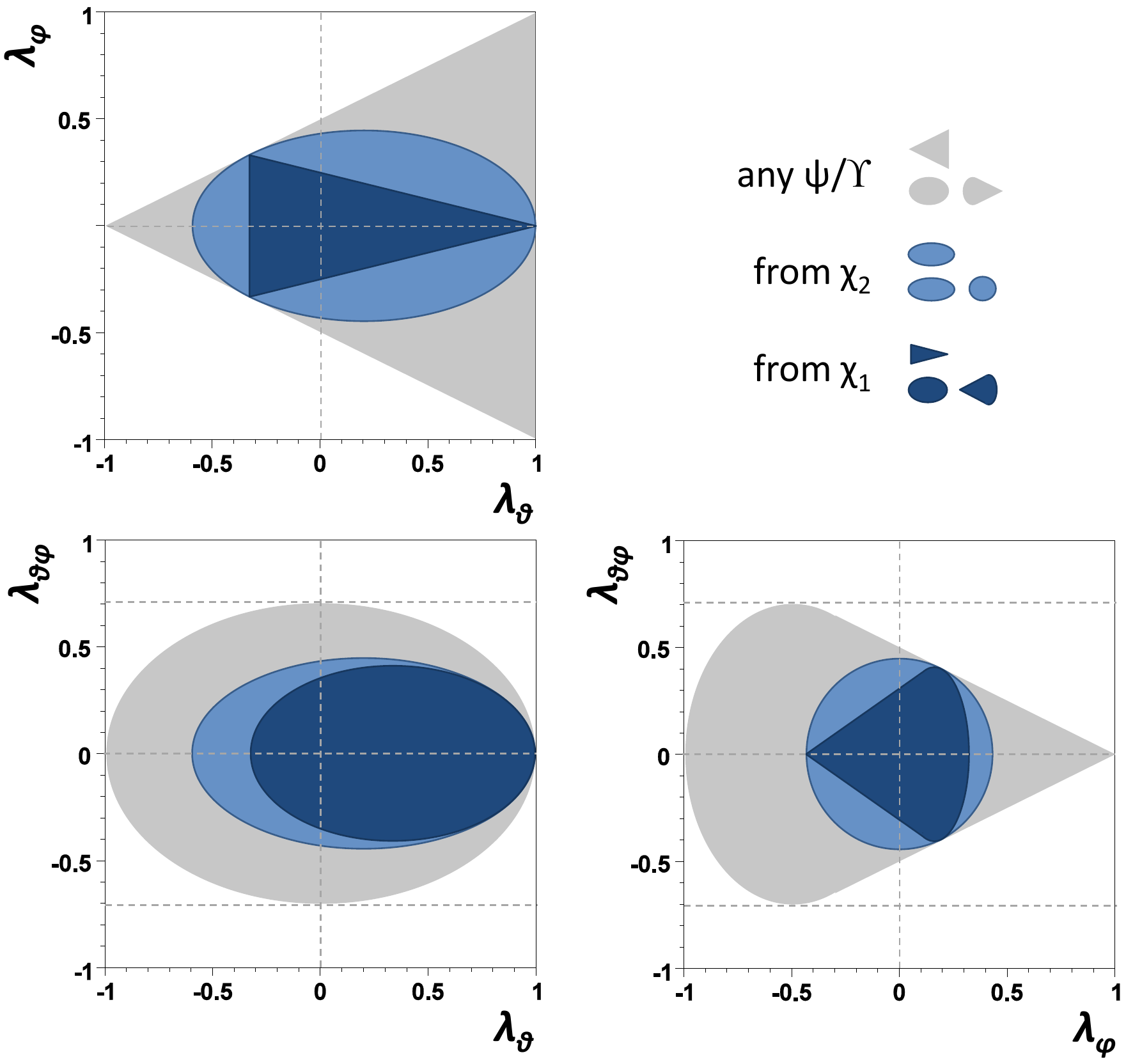}
\caption{Allowed regions for the angular parameters of the dilepton
distributions produced by the decay of vector states of any origin
(light-shaded~\cite{bib:LTviolation}), of \chiTwo\ daughters (darker)
and of \chiOne\ daughters (darkest).} \label{fig:triangles_chi}
\end{figure}

The parameters of the dilepton distribution at high momentum
(Eqs.~\ref{eq:dilepton_distr_chi1_coeffs}
and~\ref{eq:dilepton_distr_chi2_coeffs} with $\delta = 0$) satisfy characteristic
inequalities. In the \chiOne\ case,
\begin{align} \begin{split} \label{eq:triangles_chi1}
& -\frac{1}{3} \le \lambda_\vartheta \le +1 \, , \quad |\lambda_\varphi| \le
\frac{1-\lambda_\vartheta}{4} \, , \\
& \frac{9}{4}\left(\lambda_\vartheta- \frac{1}{3}\right)^2 + 6 \lambda_{\vartheta \varphi}^2 \le 1 \, ,  \\
& |\lambda_{\vartheta \varphi}| \le \frac{\sqrt{3}}{2} \left(\lambda_\varphi + \frac{1}{3}\right) \, , \\
& (6 \lambda_\varphi - 1)^2 + 6 \lambda_{\vartheta \varphi}^2 \le 1  \;\;\;
\mathrm{for} \;\;\; \lambda_\varphi > \frac{1}{9} \, .
\end{split} \end{align}
In the \chiTwo\ case,
\begin{equation} \label{eq:triangles_chi2}
\frac{5}{16}\left(\lambda_\vartheta- \frac{1}{5}\right)^2 + \lambda_\varphi^2 +
\lambda_{\vartheta \varphi}^2 \le \frac{1}{5} \, .
\end{equation}
These inequalities continue to be valid in the presence of a superposition of
production processes leading to different angular momentum compositions of the
$\chi_j$ (see the analogous discussion in Ref.~\cite{bib:LTviolation} for the
direct production of a vector state). The corresponding parameter domains are
represented in Fig.~\ref{fig:triangles_chi}, compared with the most general
constraints valid for vector states, directly or indirectly produced.
%

\section{Comment on previous calculations}
\label{sec:comment}

The angular distributions of the cascade decays $\chi_{c} \rightarrow
\mathrm{J}/\psi \, \gamma \rightarrow \ell^+ \ell^- \gamma$ were calculated in
Ref.~\cite{bib:Olsson_Suchyta} (OS) and in
Ref.~\cite{bib:Ridener_Sebastian_Grotch} (RSG) for the specific case of
low-energy $p\bar{p}$ collisions, where, due to helicity conservation, the
\chicAll\ is only produced in pure ${\rm J}_{z}$ eigenstates with eigenvalues
$m = \pm 1$ (\chicOne) or $\pm 1, 0$ (\chicTwo). The two calculations use the
$\mathrm{J}/\psi$ momentum in the \chicAll\ rest frame as quantization axis for
the dilepton, as in Fig.~\ref{fig:coords}(b), and provide the full angular
distribution of the correlated photon and lepton directions.
The result of RSG contradicts the one of OS, pointing to a seemingly wrong sign
in the last terms of the \chicTwo\ distribution (Eq.~10 of OS, corrected into
Eq.~20 of RSG) and of the \chicOne\ distribution (Eq.~15 of OS, Eq.~27 of RSG).

We checked these calculations in two ways, by repeating the steps
described in the two papers and by comparing them to our own calculation for
the full decay distribution in the special case of pure ${\rm J}_{z}$
eigenstates. In the latter case, we have applied a rotation of the lepton
variables from the $x^{\prime\prime},y^{\prime\prime},z^{\prime\prime}$ system
adopted in our calculation to the $x^{\prime},y^{\prime},z^{\prime}$ system
adopted in OS and RSG.
We found that, except for an apparent misprint of OS
(the fifth line of Eq.~11 in OS has a wrong numerical coefficient,
corrected in Eq.~21 of RSG), both calculations are correct.
RSG argued that OS
used two inconsistent conventions for the reduced rotation matrices $d^1_{ij}$,
adopting one ordering of the indices $i$ and $j$ (the one used in RSG) in the
description of the $\mathrm{J}/\psi \to \ell^+ \ell^-$ process and the reverse
ordering in the description of the $\chi_{c} \rightarrow \mathrm{J}/\psi \,
\gamma$ process. We have verified that, instead, the conventions are everywhere
consistently used, while RSG did not conform to the calculation of OS and
adopted a different definition of the photon angle.
OS refers, for the adopted notation, to Ref.~\cite{bib:Martin_Olsson_Stirling},
where the axes definitions are described in the first figure of the paper. Even
if there is no explicit mention in the text, the angle $\theta$ in the figure
(which we denote by $\Theta$ in our Fig.~1) is, unmistakably, the angle formed
by the photon momentum with the antiproton direction in the \chicAll\ rest
frame, while $\theta^\prime$ (which we denote by $\vartheta$ in our Fig.~1) is
the angle formed by the lepton momentum in the \jpsi\ rest frame with respect
to the \jpsi\ momentum in the \chicAll\ rest frame. RSG uses the same
definition of $\theta^\prime$, but an opposite definition of $\theta$: ``We
will work in the $\chi_J$ rest frame with the $Z$ axis taken to be in the
direction of $\psi$. The $\bar{p}$ direction is in the $X$-$Z$ plane, making an
angle $\theta$ with the $Z$ axis''. As a consequence, when a certain reduced
$d$ matrix is used in OS to rotate the quantization axis by an angle $\theta$,
the inverse rotation must appear in the calculation of RSG. If
$d^1_{ij}(\theta)$ represents a given rotation, the inverse rotation can be
denoted either by exchanging $i$ with $j$ (this induced RSG's misinterpretation
of the discrepancy) or by replacing $\theta$ with $2\pi-\theta$. This explains
the different sign in the term proportional to $\sin2\theta$ resulting from the
two calculations. The remaining terms, depending on $\cos^2\!\theta$, are not
sensitive to such a redefinition of the angle.

In short, each of the two calculations is correct, if they are made with the
matching angle definition. If, on the contrary, the definition of $\theta$ used
by OS is used together with the distributions functions derived in RSG, or
vice-versa, a wrong sign appears in the term proportional to $\sin2\theta$,
leading to \emph{unphysical} results. In fact, this artificial change of sign
is not reabsorbed in a different definition of sign and/or magnitude of the
higher-order multipole amplitudes: already in the E1 approximation, the
physical correlation between photon and lepton angles is \emph{substantially}
altered by such a mistake. To evaluate the importance of this problem, we
assumed the angle definitions of OS and used the formulas derived in RSG,
transposing them, by rotation, to the system of axes used in our calculations
[Fig.~\ref{fig:coords}(c)]. As a result of this forced mistake, we arrive to a
physical result which is almost opposite to the correct one: the lepton
distribution, instead of being a perfect clone of the photon distribution (in
the E1 limit), becomes a consistently smeared, almost isotropic distribution,
for whatever polarization state of the \chiAll\ (in other words, the domains of
the \chiOne\ and \chiTwo\ dilepton parameters, represented in
Fig.~\ref{fig:triangles_chi}, are reduced to small areas around the origin).

We have noticed that the measurements of E760~\cite{bib:E760} and
E835~\cite{bib:E835}, included in the present world averages of $h_2$, $g_2$
and $g_3$ in the Review of Particle Physics~\cite{bib:PDG}, seem to be affected
by this kind of misunderstanding. Both analyses define the photon angle
$\theta$ as ``the polar angle of the \jpsi\ with respect to the antiproton'',
as in OS, but the formulas are taken from RSG (Table~II in the E760 paper and
Tables~IV--V in the E835 paper reproduce Eqs.~20 and~27 of RSG). On the other
hand, the quality of the global fits of the data using the adopted
parameterization is rather good and the measurements of the higher-order
multipoles are compatible with the CLEO results~\cite{bib:CLEO}, suggesting
that the inconsistency between formulas and angle definitions might simply be
an editing mistake in both experimental papers.

\section{Conclusions}
\label{sec:conclusion}

We have derived the expressions of the angular distributions of the radiative
decay from a $^3P_J$ state to a $^3S_1$ state and of the dilepton decay of the
latter. No selection rules specific to certain quarkonium production mechanisms
have been used and the choice of the polarization frame for the directly
produced states has been kept completely general.

We have shown that the \chiAll\ polarizations can be measured (for
not-too-low-momentum experiments) directly from the angular distribution of the
\emph{dilepton} decay in the \jpsi~/~\upsAll\ rest frame, with respect to the
same kind of system of axes (Collins--Soper, helicity, etc.) adopted in
inclusive \jpsi~/~\upsAll\ measurements.

In fact, the dilepton distribution in the \jpsi~/~\upsAll\ rest frame is a
clone of the photon distribution in the \chiAll\ rest frame, \emph{stripped of
the contribution of the higher multipoles of photon radiation}.

This represents a significant advantage, given that such contributions ---
measured to be quite important in the \chicTwo\ case, poorly known (due to
contradictory measurements) in the \chicOne\ case, and still unmeasured for the
bottomonium family --- can have a very large impact in the measurement.
Furthermore, a simultaneous determination of \chiAll\ polarization \emph{and}
of the multipole parameters is scarcely feasible at hadron colliders.

An additional advantage of this method is that it does not use the photon
measurement to reconstruct the event-by-event decay topology. 
This means that, contrary to previous expectations, the measurement of \chiAll\ 
polarization is not intrinsically more challenging than, for instance, the 
measurement of the \chicOne/\chicTwo\ cross section ratio (a measurement 
presently being done by several experiments at the Large Hadron Collider).
In both cases the analysis needs to identify an event sample where the \jpsi\ 
(or $\Upsilon$) dilepton is associated to a photon giving an invariant mass 
of the $\mu^+\mu^-\gamma$ system in the \chiAll\ mass region.  This is usually
done using photons reconstructed by the conversion method, given that the
tracking of the electron-positron pair gives good enough resolutions to resolve
the \chicOne\ and \chicTwo\ resonances.
Naturally, a larger event sample is needed for a multi-dimensional angular analysis. 
But there are no extra difficulties related to photon backgrounds or reconstruction 
efficiencies depending on the decay angles, specific to the measurement of the 
polarization (as would be the case using the previously available methods).

It is worth reminding that a certain numerical value of the observable
polarization parameters corresponds to very different quantum-mechanical states
of the \chiOne\ and \chiTwo\ (e.g., $\lambda_{\vartheta}=+1$ can reflect the
$J_z = 0$ state of the \chiOne\ or the $J_z = \pm 2$ state of the \chiTwo).
Therefore, a reliable experimental discrimination between the \jpsi\ or
\upsAll\ coming from the decays of these two states is crucial for a proper
understanding of \chiAll\ polarization.

We have also pointed out misunderstandings in previous calculations, which may
have affected some of the existing measurements of the higher-order photon
multipoles in \chicAll\ decays.

\bigskip

  P.F., J.S.\ and H.K.W.\ acknowledge support from Funda\c{c}\~ao para
  a Ci\^encia e a Tecnologia, Portugal, under contracts
  SFRH/BPD/42343/2007, CERN/FP/109343/2009 and SFRH/BPD/42138/2007.


\end{document}